\documentclass{mn2e}
\let\bibhang\relax
\usepackage{graphicx}
\usepackage{txfonts}
\usepackage{natbib}
\bibpunct{(}{)}{;}{a}{}{,}

\title[Gas and dust around A-stars at tens of Myr]
{Gas and dust around A-type stars at tens of Myr: signatures of 
cometary break-up
}
\author[J. S. Greaves et al.]{J. S. 
Greaves$^1$\thanks{E-mail: GreavesJ1 at cardiff.ac.uk}, W.S. Holland$^2$, 
B.C. Matthews$^3$, 
J.P. Marshall$^4$, W.R.F. Dent$^5$, 
\newauthor{P. Woitke$^5$, M.C. Wyatt$^7$, 
L. Matr\`{a}$^7$ \& A. Jackson$^8$ }\\
$^1$School of Physics \& Astronomy, Cardiff University, 4 The Parade, Cardiff CF24 3AA, UK\\
$^2$Astronomy Technology Centre, Royal Observatory Edinburgh, Blackford Hill, Edinburgh EH9 3HJ, UK \\
$^3$National Research Council of Canada, 5071 West Saanich Rd, Victoria, BC, V9E 2E7, Canada \\
$^4$School of Physics, University of New South Wales, NSW, 2052, Sydney, Australia \\
$^5$ALMA Santiago Central Offices, Alonso de Córdova 3107, Vitacura, Casilla 763 0355, Santiago, Chile \\
$^6$School of Physics \& Astronomy, University of St Andrews, North Haugh, St Andrews, Fife KY16 9SS, UK \\
$^7$Institute of Astronomy, Madingley Road, Cambridge CB3 0HA, UK \\
$^8$School of Earth and Space Exploration, Arizona State University, Tempe, AZ 85287, USA
}

\begin{document}

\date{Accepted 2016. Received 2016; in original form 2016}

\pagerange{\pageref{firstpage}--\pageref{lastpage}} \pubyear{2016}

\maketitle

\label{firstpage}

\begin{abstract}

Discs of dusty debris around main-sequence stars indicate fragmentation of 
orbiting planetesimals, and for a few A-type stars, a gas component is also 
seen that may come from collisionally-released volatiles. Here we find the 
sixth example of a CO-hosting disc, around the $\sim$30 Myr-old A0-star HD 
32997. Two more of these CO-hosting stars, HD 21997 and 49 Cet, have also 
been imaged in dust with SCUBA-2 within the SONS project. A census of 27 
A-type debris hosts within 125 pc now shows 7/16 detections of 
carbon-bearing gas within the 5-50 Myr epoch, with no detections in 11 
older systems. Such a prolonged period of high fragmentation rates 
corresponds quite well to the epoch when most of the Earth was assembled 
from planetesimal collisions. Recent models propose that collisional 
products can be spatially asymmetric if they originate at one location in 
the disc, with CO particularly exhibiting this behaviour as it can 
photodissociate in less than an orbital period. Of the six CO-hosting 
systems, only $\beta$ Pic is in clear support of this hypothesis. However, 
radiative transfer modelling with the ProDiMo code shows that the CO is 
also hard to explain in a proto-planetary disc context.

\end{abstract}


\section{Introduction}

Dusty debris around main-sequence stars results from collisions between 
rocky bodies. Timescales for the particles to fall into the star or grind 
down to sizes small enough to be blown out by radiation pressure are short 
compared to stellar lifetimes, so larger progenitor planetesimals must be 
present. In most cases, these are found to be in belts located at tens of 
AU, where rock/ice comet-like compositions are probable, akin to solar 
system cometary material. Although collisions should cause the ices to 
sublimate into gases, this component in the belts is difficult to detect, 
as molecules are quickly photo-dissociated. For nearby debris-hosting 
stars, e.g. Fomalhaut, improved limits from deep observations are 
important for comparing the chemistry to solar system comets (Matr\`{a} et 
al. 2015).

There are five A-type debris hosts where the molecule carbon monoxide (CO) 
has been detected, via millimetre rotational transitions. This species 
photo-dissociates on hundreds of year timescales even under interstellar 
radiation (Zuckerman \& Song 2012, e.g.). When illuminated by A-stars, 
this timescale can be shortened to less than orbital periods at tens of 
AU (Jackson et al. 2014, e.g.). If CO is preferentially produced at one 
location in the disk then the CO distribution would be `one-sided' 
around the star, and this effect has been imaged recently in an ALMA 
study of $\beta$ Pic (Dent et al. 2014; Matr\`{a} et al., in prep.).

Other gas phases can also be present, with CII and OI lines in the 
far-infrared (Dent et al. 2012) potentially tracing photo-dissociated CO 
(Roberge et al. 2013). Three of the CO-hosting systems also appear to host 
`falling evaporating bodies', with transient red-shifted Ca-absorption 
features seen towards $\beta$ Pic and 49 Cet (Montgomery \& Welsh 2012), 
and Na absorption identified towards HD 32297 (Redfield 2007). These 
features are consistent with volatiles originating from ongoing cometary 
breakups, an idea which is now being explored by models (Kral et al. 2016, 
e.g.).

Here we consider debris-hosting A-stars within 125 pc (parallax 
$\geq$8 mas) that have been searched for CO. This distance limit helps 
to exclude stars outside the Local Bubble where interstellar CO may be 
a strong contaminant. We report the sixth detection of CO, around the 
approximately 30 Myr-old A0 star HD 32297. In this case, we 
successfully used the presence of weak CII emission (Donaldson et al. 
2013) as a predictor for the presence of CO.

We have also followed CO detections for 49 Ceti and HD 21997 (Zuckerman 
et al. 1995; Mo\'{o}r et al. 2011) with dust continuum-imaging at 450 and 
850 micron wavelength. These continuum data are part of the JCMT Legacy 
Project SONS (SCUBA-2 Survey of Nearby Stars), described by Pani\'{c} et 
al. (2013). The gas-plus-dust systems are also tested here against model 
predictions that the short-lived CO component could be more spatially 
asymmetric than the dust.

\begin{figure}
\label{fig1}
\includegraphics[width=0.5\textwidth,angle=0,trim=0 3cm 0 3cm,clip]{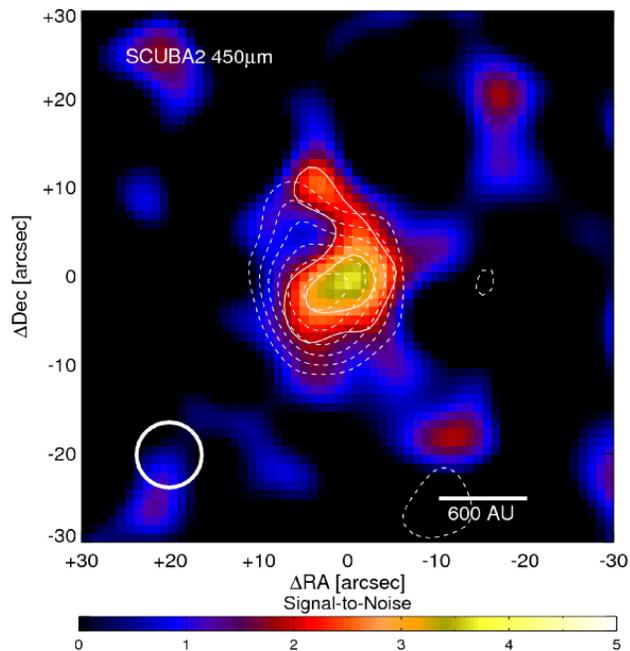}
\caption{SCUBA-2 results for 49 Cet, on signal-to-noise ratio scales. 
The peak flux at 450 microns is is 74 $\pm$ 14 mJy/beam, and the 
integrated flux within a 40 arcsec diameter aperture is 125 $\pm$ 10 
mJy. The dashed contours show the 850~$\umu$m SNR (peak = 8.3), overlaid 
on the 450~$\umu$m colour-scale. The integrated flux at 850~$\umu$m is 12.1 
$\pm$ 2.0 mJy. The secondary peak adjacent to the north (top) side of the disc 
may be due to a background dusty galaxy. The stellar position coincides with 
the 450~$\umu$m flux-peak within typical pointing drifts ($\la2''$).
}
\end{figure}

\begin{figure}
\label{fig2}
\includegraphics[width=0.5\textwidth,angle=0,trim=0 3cm 0 3cm,clip]{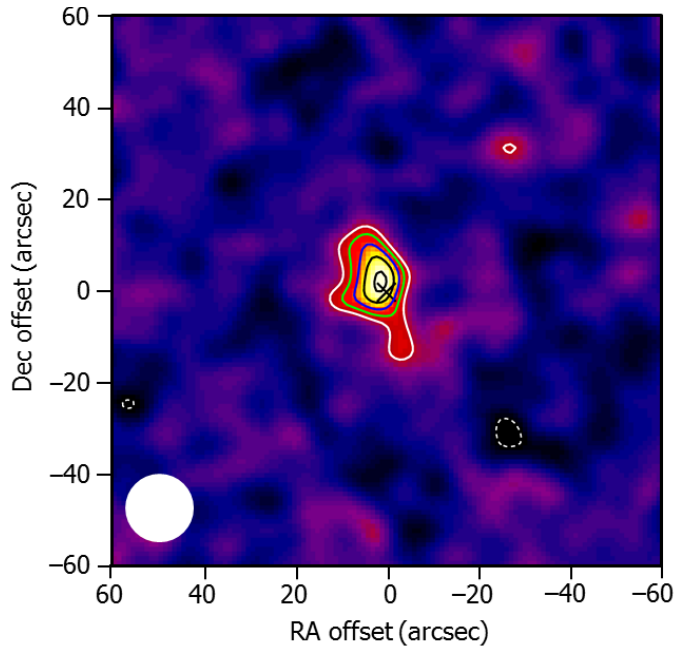}
\caption{SCUBA-2 850 micron image for HD 21997, with contours at 3,4,5,6,7 sigma 
levels and a peak of 7.9 $\pm$ 1.1 mJy/beam. The integrated flux is 10.7 $\pm$ 
1.5 mJy. 
}
\end{figure}

\section{Observations}

The new data were obtained with the 15 m James Clerk Maxwell Telescope 
located on Mauna Kea, Hawaii. Observing procedures have been described 
by, for example, Pani\'{c} et al. (2013, 2010). HD 21997 and 49 Cet 
were observed with the SCUBA-2 camera (Holland et al. 2013) in 2012-2015, 
in moderately dry conditions (225~GHz zenith opacities under 0.1). 
Images are shown with 1 arcsec pixels, and after smoothing with a 
7-arcsec Gaussian, the effective beam diameters are 15.8 and 11.6 
arcsec at 850 and 450 microns respectively. The continuum data 
reduction used the makemap task in the SMURF package (Jenness et al. 
2011), plus high-pass filtering and zero-masking to remove residual 
low-frequency structure in the backgrounds.

The CO data for HD 32297 at 1.3 mm were taken as part of a poor weather 
backup programme in 2013 (225~GHz zenith opacities up to 0.3), using the 
RxA3 receiver to search for the J=2--1 transition. The spectral reduction 
used the makecube task in SMURF plus the SPLAT package for baselining and 
binning. A 200 km/s velocity range is shown around the stellar velocity in 
the heliocentric frame of 23.0 $\pm$ 0.3 km/s (Torres et al. 2006). 
After about 10 hours on sky, the JCMT J=2-1 data are about twice as 
sensitive as the corresponding spectrum of Zuckerman \& Song (2012) from 
the IRAM 30m telescope.

\section{Results}

\subsection{Dust imaging}

\begin{figure}
\label{fig3}
\includegraphics[width=35mm,angle=90,trim=4cm 0 4cm 0,clip]{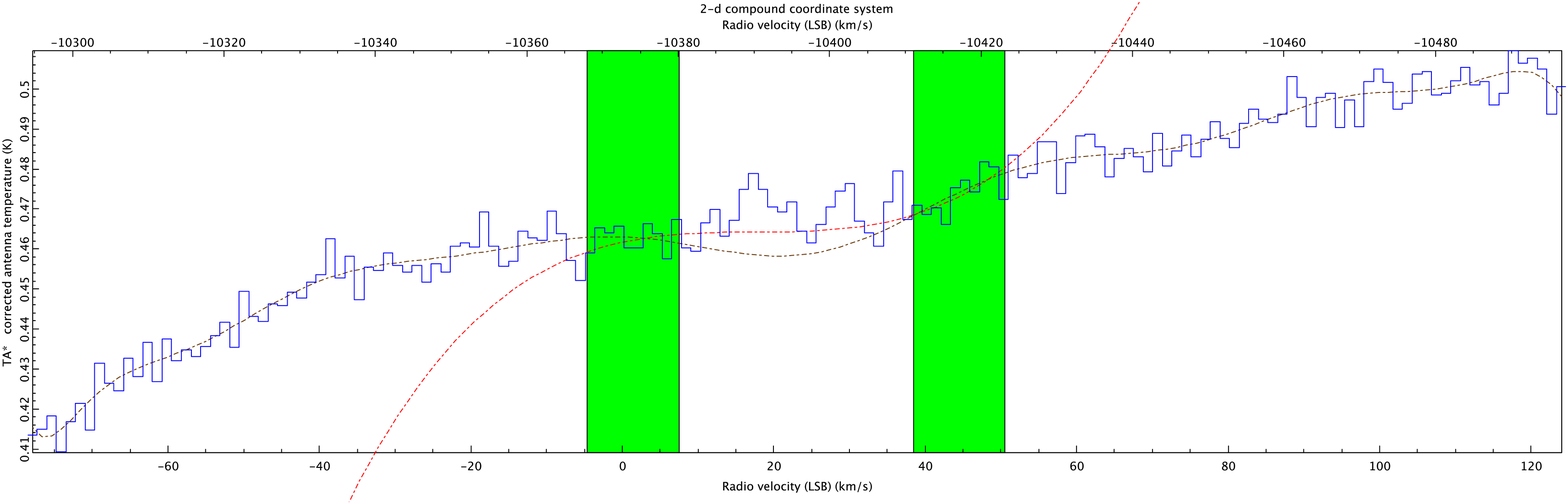}
\includegraphics[width=50mm,angle=90,trim=2cm 0 2cm 0,clip]{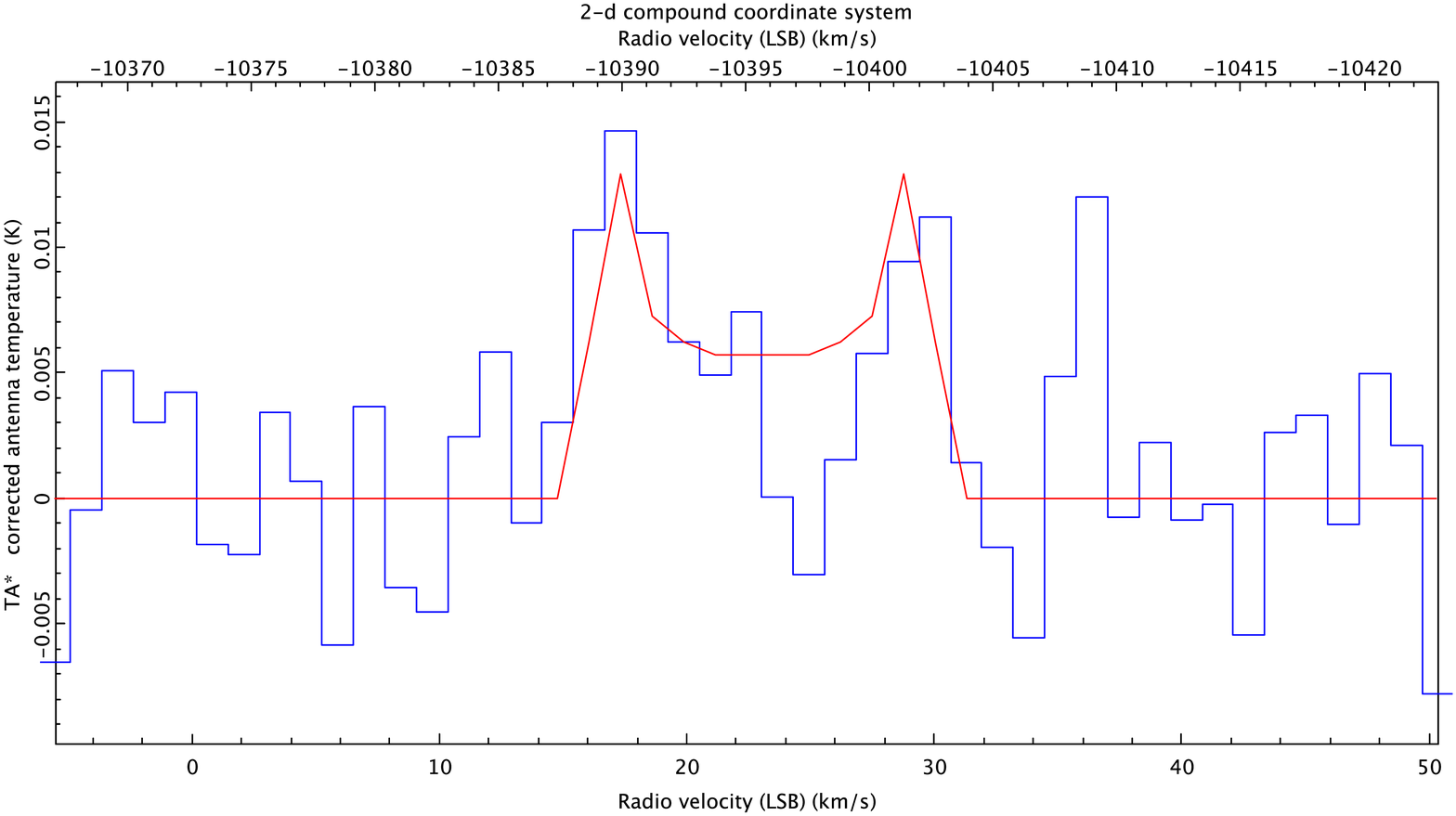}
\includegraphics[width=50mm,angle=90,trim=2cm 0 2cm 0,clip]{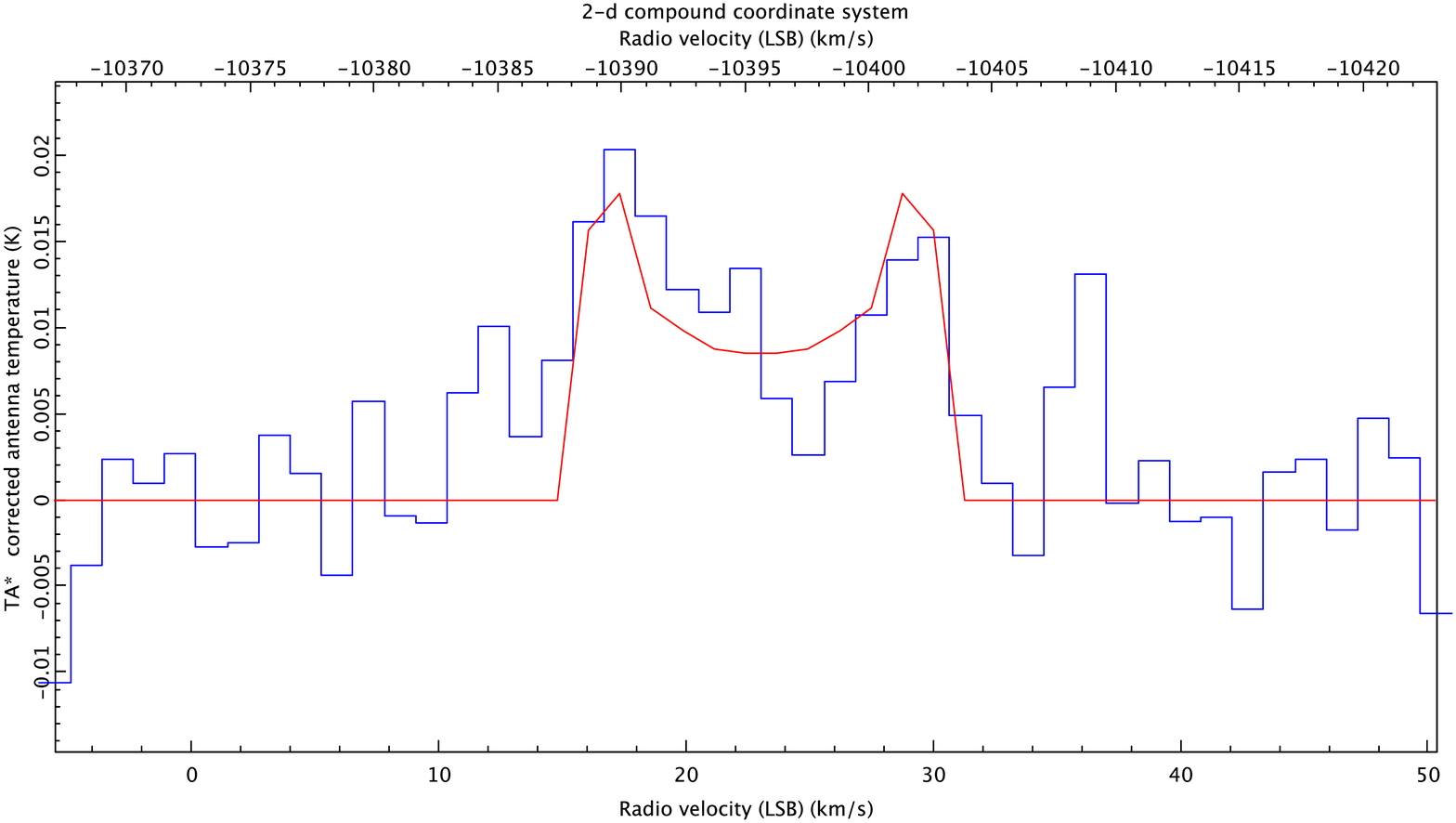}
\caption{CO J=2-1 spectrum for HD 32297. Top: 
results from fitting the passband 
with wide and narrow (green-shaded) velocity ranges. 
Middle: 
spectrum after subtracting the 3rd-order polynomial shown in red in the top 
panel. 
Bottom: spectrum after 
subtracting the 14th-order polynomial shown in brown in the top panel. 
The two models shown by the red curves in the middle and bottom panels are 
for molecules on Keplerian 
orbits (see text). 
}
\end{figure}

Figure 1 shows the SCUBA-2 data for 49~Cet. The disc is spatially resolved 
at 450~$\umu$m, at the expected orientation. For comparison, the position 
angle with Herschel at 70~$\umu$m (Roberge et al. 2013) is $\approx 
105^{\circ}$ (anti-clockwise from north), while CO lies at PA $= 110 \pm 
10^{\circ}$ (Hughes et al. 2008). The SExtractor tool for automatic object 
detection\footnote{ The peak just to the north was assumed to be unrelated, 
and was blanked before fitting the disc.} was used here, indicating PA 
$\approx120^{\circ}$ for the 450~$\mu$m disc axis.

At the half-power contour at 450~$\mu$m, the major:minor axis ratio is 
1.52 from this fit. The long diameter is 17.7~arcsec and the minor axis 
is unresolved. This is consistent with the disc being seen close to 
edge-on -- Lieman-Sifry \& Hughes (2015) find an inclination of $79.6 
\pm 0.4^{\circ}$ from their ALMA 850~$\umu$m image. They also identify 
dust out to $286 \pm 7$ AU from the star, while our 450~$\umu$m image 
(after deconvolving the beam size in quadrature) gives a slightly larger 
outer radius, approximately 6.7~arcsec or 395~AU at 59~pc.

Figure 2 shows our 850~$\umu$m image of HD 21997. The structure appear 
asymmetric, with most of the flux centred in a peak slightly north of the 
star and only a 3-sigma counterpart to the south. In contrast, Mo\'{o}r et 
al. (2013) found a compact symmetric disc in their ALMA 886 $\mu$m image. 
However, they noted some extension along an axis at PA $\sim 20-25^{\circ}$ 
in Herschel images at 70-100 $\mu$m. This agrees with the slightly 
east-of-north orientation seen in the SCUBA-2 image. To reconcile the 
submillimetre datasets, there could be significant flux beyond the 
half-power point of the ALMA primary beam ($>$9 arcsec, or 650 AU). The 
SCUBA-2 integrated flux in fact substantially exceeds the 2.7 $\pm$ 0.3 mJy 
in a 6-arcsec aperture from ALMA. Only 15 ALMA antennas were available to 
Mo\'{o}r et al., so some spatial scales may not have been fully sampled. 

SCUBA-2 also simultaneously observed HD 21997 at 450 $\mu$m, and a 
tentative signal of 85 $\pm$ 25 mJy/beam was seen near the star. There is a 
similar peak about 15 arcsec SW of the star, but this appears to coincide 
with a faint part of the 850 $\mu$m background structure.

%Tests of image fidelity have been made on the SONS dataset, by shifting 
%individual SCUBA-2 observations to correct for drifts recorded on 
%pointing sources. Such drifts were found to be typically 1.5 arcsec at 
%the mid-observation point, and the corrections made no significant 
%changes to the co-added images. We conclude that the offset of the 
%large-scale flux peak is real for HD 21997, which has no unusual 
%pointing history. There is also a more distant 3-sigma peak on the 
%opposite side of the star, which could be fitted as part of a large 
%non-symmetric disc. However, at the depth of the SONS 850~$\mu$m data, 
%there is a $\approx$1-in-25 probability that a beam this far from the 
%star contains an unrelated background object.

HD 32297 is not a SONS target, but has been imaged in dust emission at 
1.3 mm by Manness et al. (2008), and was resolved at arcsec resolution. 
The disc was found to be extended and elliptical, with a strong 
asymmetry, in that the centroid is offset from the star with 4-sigma 
confidence.

\subsection{CO spectroscopy} 

\begin{table*}
 \caption{Literature data for the 27 debris-hosting A-stars within 125 pc 
that have been searched for CO, sub-divided into two age groups. The CII 
and OI lines were searched for with {\it Herschel}; the Herschel Science 
Archive was used to check a few unpublished spectra. Blank entries indicate 
no observation has been made, while u.l. denotes an upper limit. Stellar 
ages are from 
(1) Manoj et al. (2006); 
(2) TWA association: Ducourant et al. (2014); 
(3) LCC association: Song et al. (2012); 
(4) Rhee et al. (2007); 
(5) Beta Pic Moving Group: Mamajek \& Bell (2014); 
(6) Kalas (2005); 
(7) Argus association: Zuckerman \& Song (2012); 
(8) Tuc-Hor moving group: Bell et al. (2015); 
%(9) Rieke et al. (2005);
(9) Chen et al. (2014); 
(10) Mamajek (2012); 
(11) McDonald et al. (2012);
(12) Monnier et al. (2012);  
(M) Mo\'{o}r et al. (2015,2006).
}
 \label{tab:list}
 \begin{tabular}{@{}lllrrcccr}
  \hline
star	& alias	& spectral	& distance & age  & CO? & CII? & OI? 	& {\em incidences in age group} \\
	&	& type	  	& (pc)	   & (Myr)&     &      & 	& \\
  \hline
%HD35187	&	& A2e+	&114	& 6 (1)		&u.l.	&u.l.	&Yes	\\
HD141569& 	& A0Ve	&99	& 7 (1) 	&Yes	&Yes	&Yes	\\
HD109573&HR4796	& A0	&73	& 8 (2)		&u.l.	&	&	\\
%HD100453& 	& A9Ve	&122	& 10 (3)	&u.l.	&u.l.	&Yes	\\
HD110058& 	& A0V	&107	& 10 (4)	&u.l.	&	&	\\
HD131835&	& A2IV	&123	& 16 (M)	&Yes	&u.l.	&u.l.	\\
HD95086	&	& A8III	&90	& 17 (M)	&u.l.	&	&	\\
HD121617&	& A1V	&120	& 17 (M)	&u.l.	&	&	\\
HD39060 &betaPic& A6V	&19	& 23 (5)	&Yes	&Yes	&Yes	\\
HD172555& 	& A7V	&29	& 23 (5)	&u.l.	&u.l.	&Yes	\\
HD181296&etaTel	& A0V	&48	& 23 (5)	&u.l.	&Yes	&u.l.	\\
HD32297 &	& A0V	&112	& 30 (6)	&Yes	&Yes	&u.l.	\\
HD38206	&	& A0V	&75	& 30 (M)	&u.l.	&	&u.l.	\\
HD85672	&	& A0	&93	& 30 (4)	&u.l.	&	&	\\
HD9672 	&49Cet	& A1V	&59	& 40 (7)	&Yes	&Yes	&u.l.	\\
HD3003 	&beta3Tuc& A0V	&46	& 45 (8)	&u.l.	&	&u.l.	\\
HD21997 &HR1082	& A3IV/V&72	& 45 (8)	&Yes	&u.l.	&u.l.	\\
HD102647&betaLeo& A3Va	&11	& 50 (4)	&u.l.	&	&	\\
&&&&&&&& {\em CO and/or C$^+$: 7/16 (44\%); OI: 3/10 (30\%) }	\\
\hline
HD110411&	& A0V	&36	& 90 (M)	&u.l.   &       &       \\	
HD17848 &nuHor	& A2V	&51	& 100 (4)	&u.l.	&	&	\\
%HD182681&	& B8/9V &69.9	& 140 (M)	&u.l.   &       &       \\
HD183324&	& A0V	&61	& 140 (M)	&u.l.   &       &       \\
HD182919&	& A0V	&73	& 200 (M)	&u.l.   &       &       \\
HD161868&gammaOph&A1V..	&32	& 200 (4)	&u.l.	&u.l.	&	\\	
%HD164577&68Oph	& A2Vn	&89.7	& 260 (9)	&u.l.	&	&	\\
HD95418	&	& A1V	&24	& 320 (9)	&u.l.	&	&	\\
HD10939	&	& A1V	&62	& 350 (M)	&u.l.	&	&	\\
HD216956&Fomalhaut& A4V	&7.7	& 440 (10)	&u.l.	&u.l.	&u.l.	\\
HD6028	&	& A3V	&91	& 500 (11)	&u.l.	&	&	\\
%HD81515 &	& A5Vm	&117.2	& 530 (11)	&u.l.	&	&	\\
HD158352&HR6507	& A7V	&60	& 600 (4)	&u.l.	&u.l.	&u.l.	\\
HD172167&Vega	& A0Va	&7.7	& 700 (12)	&u.l.	&u.l.	&u.l.	\\
&&&&&&&& {\em CO and/or C$^+$: 0/11 (0\%); OI: 0/3 (0\%)	} \\
\hline
 \end{tabular}
\end{table*}

Figure 3 shows the CO spectrum towards HD 32297, reduced by two different 
methods to test for robustness. The entire passband was first fitted with a 
high-order polynomial, in order to fully use the information present, and 
excluding only a region of 20 channels of 1.27 km/s around the stellar 
velocity (+23.0 km/s, Torres et al. 2006). Secondly, a least-parameters 
approach was used, fitting a low-order polynomial only across the two 
green-shaded velocity ranges. The noise residual was 3.75 mK per channel in 
the latter case, and 4.00 mK in the former (this fit was possibly limited 
by the maximum 14th-order fit available in the SPLAT software). Subtracting 
the narrow-fit baseline results in a detection with 6.1 sigma confidence, 
and integrated line signal of 106.4 $\pm$ 17.5 mK km/s. The wide-fit 
baseline yields a 10.5 sigma detection of 200.0 $\pm$ 19.1 mK km/s. Both 
velocity-intervals were 14 channels wide. In the latter case, the remainder 
of the baselined spectrum could be used to check for false-positives, and 
the maximum integrated signal found in a sliding 14-channel-wide window was 
17.1 mK km/s. The intensities are in a T$_A^*$ antenna brightness 
temperature scale\footnote{The Jy/K conversion factor for the JCMT is 
15.6/$\eta_a$, with aperture efficiency $\eta_a$ having standard values of 
0.52 and 0.61 at 345 and 230 GHz respectively.}.

The preferred spectrum results from the wide-fit baseline subtraction, 
because the depth of the central minimum is more consistent with gas 
molecules on Keplerian orbits. Because of velocity projection effects, the 
central minimum should always have a positive signal, even for an 
infinitesimally-thin ring. The toy models shown adopt such a ring at 44 AU 
radius for the narrow-fit case, compared to a belt at 35-45 AU in the 
wide-fit case. The former model over-predicts the data by 2.4 sigma in the 
worst-case channel, and the latter by only 1.5 sigma. Also, the former case 
has a residual (summed variance between model and data) of 207 mK$^2$ 
compared to 164 mK$^2$ expected from random noise, while the latter fits 
within the noise (variance of 137 mK$^2$ compared to random expectation of 
192 mK$^2$). The models are only intended to be indicative, but are 
informative as to possible locations of the gas orbits. For our fits to 
velocities in an edge-on disc around a 2.1 M$_{\odot}$ host star 
(Boccaletti et al. 2012), the radii thus point to molecules lying inwards of 
the dust -- for example, Donaldson et al. (2013) have fitted a dust ring 
with a radius of around 110 AU.

There is a possible asymmetry between the red and blue sides of the line 
profile, which in the preferred fit have integrated intensities of 76.3 and 
123.7 mK km/s respectively. The error on the difference is $\sqrt{2} \times 13.4$ mK 
km/s, so this has only 2.5 sigma confidence. (In the narrow-fit case, the 
respective quantities are 33.4 and 73.0 mK km/s, differing at the 2.25 sigma 
level.)

A CO J=3-2 line towards HD 32297 is tentatively seen in a 
short JCMT observation with the HARP camera, from the archive. 
The integrated signal is 226 $\pm$ 92 mK km/s (2.5 sigma) over a  
$v_{star} \pm 9$ km/s velocity range. However, Mo\'{o}r et al. (2011) 
reported a CO 3-2 upper limit from APEX which would translate to $\la$150 
mK km/s on the JCMT T$_A^*$ scale, for nominal telescope efficiencies. 

The mean intensities across the JCMT 2-1 and 3-2 lines are 11.3 $\pm$ 1.1 
mK (wide-fit baseline case) and 12.7 $\pm$ 5.2 mK, with the latter value 
probably $\la$8.4 mK for consistency with the APEX data. This gives a 
3-2/2-1 line ratio of $\sim$0.50 or less (after correction for the 
different beam-filling factors for a presumed point-like disc). In this 
case, the gas excitation temperature would be very low, at below $\sim$5 K, 
but this only applies in a regime where both lines are taken to be 
optically thin. This is not true in some other systems, as noted below; see 
also Matr\`{a} et al. (2015) for further discussion of excitation of CO 
around an A-star.

\subsection{CO census}

From a master list of debris systems searched for CO rotational 
transitions, we find 27 A-star hosts within 125 pc of the Sun (Table 1). 
The most recent systematic survey is that of Mo\'{o}r et al. (2015), who 
also summarise previous work. The combined samples are close to complete 
for fractional dust luminosities $L_{dust}/L_* > 10^{-4}$, and have 
particularly targeted objects up to ages of 50 Myr. Our new target HD 32297 
is the sixth system where a CO line has been detected. The earlier 
archetypes were found by Zuckerman et al. (1995), using the IRAM 30 m 
telescope to discover CO 2-1 emission around HD 145169 and 49 Ceti. 
Corresponding CO 3-2 lines were found with the JCMT (Dent et al. 1995, 
2005). Mo\'{o}r et al. (2011, 2015) have subsequently discovered CO 
transitions in the HD 21997 and HD 131835 systems using the APEX 12m 
telescope. Dent et al. (2014) have been the first to map a distribution of 
CO, observing the 3-2 transition in the $\beta$ Pic disc.

\subsection{Properties of discs with gas}

We now examine the seven C-hosting A-star systems, which have ages ranging 
from around 7 to 45~Myr (Table 1). This group is unified by the presence of 
CO and/or C$^+$ -- and often both signatures, in support of the hypothesis 
that C$^+$ may be a by-product of CO photo-dissociation. (Observational 
limits may prevent this being universally seen: HD 21997 and HD 131835 have 
CO but only upper limits on CII, while $\eta$ Tel shows CII but not CO.) In 
contrast, OI persists for only about half as long, with the latest 
detection being in the 23 Myr Beta Pic Moving Group (BPMG). Strong OI 
emission is a characteristic of proto-planetary discs that are less evolved 
than debris discs (Dent et al. 2013).

\begin{figure}
\label{fig4}
\includegraphics[width=0.37\textwidth,angle=90,trim=2cm 2cm 2cm 4cm,clip]{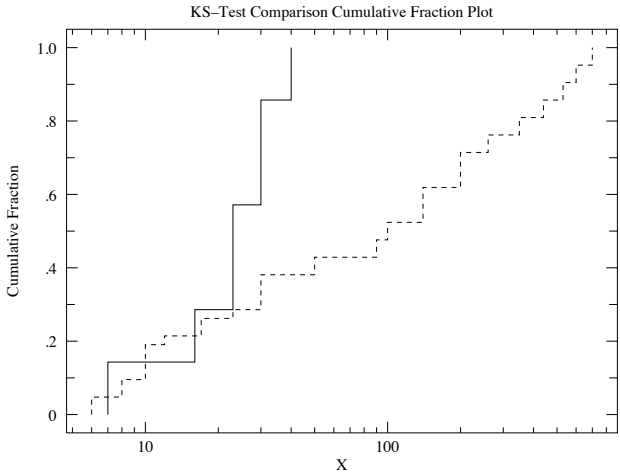}
\caption{Cumulative distributions for debris discs with carbon gas 
(solid line) and without (dashed line), as functions of the stellar 
age in Myr. 
}
\end{figure}

\begin{table*}
 \caption{Continuum and line fluxes for the A-stars with CO detections, plus 
the CII-detected system $\eta$ Tel. The 850-870 micron fluxes are from 
observations here and in the literature, except for HD 32297 (estimated by 
interpolation over 500--1200 micron: Donaldson et al. 2013). The {\it 
Herschel}-detected lines of CII and OI are at wavelengths of 158 and 
63 microns respectively. All line fluxes are in units of 10$^{-18}$ W/m$^2$; 
continuum fluxes are in mJy (10$^{-29}$ W/m$^2$/Hz). Uncertainties in brackets 
are 1-sigma and upper limits are 3-sigma.}
 \label{tab:flux}
 \begin{tabular}{@{}llllllll}
  \hline
	   & HD 141569      & HD 131835		& beta Pic		& $\eta$ Tel         & HD 32297      		& 49 Cet	   & HD 21997 \\
 \hline
$L_{dust}/L_*$ & 7e-3 	    & 3e-3		& 3e-3	      		& 2e-4               & 3e-3	        	& 1e-3	  	   & 5e-4	\\
F(850/870) & 12.6($\pm$4.6) & 8.5($\pm$4.4)	& 104($\pm$10)  	& $\leq$ 14.4        & 9($\pm$3)         	& 14.8($\pm$3.1)   & 6.3($\pm$1.6) \\
F(CO 3-2)  & 0.14($\pm$0.1) & 0.032($\pm$0.006)	& 0.076($\pm$0.008)  	& $\leq$ 0.083       & 0.065($\pm$0.027) 	& 0.069($\pm$0.014)& 0.038($\pm$0.006) \\
F(CO 2-1)  & 0.03(---) 	    & $\leq$0.018	& $\leq$ 0.03  		&                    & 0.037($\pm$0.004) 	& 0.013(---)  	   & 0.018($\pm$0.004) \\
F(CII)     & 11.4($\pm$1.8) & $\leq$0.5		& 24.3($\pm$0.4)        & 1.7($\pm$0.4)      & 2.68($\pm$0.72)   	& 0.80($\pm$0.17)  & $\la$ 1.5 \\
F(OI)	   & 245($\pm$5)    & $\leq$1.5		& $\approx$ 7  	        &                    & $\leq$ 7.3        	& $\leq$ 11   	   & $\la$ 2.7\\
  \hline
 \end{tabular}
\end{table*}

Figure 4 shows cumulative distribution functions for the systems with and 
without carbon gas, as a function of the stellar age. At around 30 Myr, 
there is a divergence in the two populations, with carbon-bearing systems 
ceasing by 50 Myr, but two-thirds of the gas-less systems having greater 
ages. There do not seem to be strong selection effects, as the list of 
systems from Table~1 is spread quite evenly in log-age, reflecting a rather 
typical distribution for detectable solid debris. The age distribution is 
not intrinsically that of A-type stars, because debris fades with time (at 
a range of evolutionary rates). Further, CO searches have concentrated on 
systems with $L_{dust}/L_* > 10^{-4}$, which could introduce skews related 
to both evolution and detectability by distance -- however, any biases here 
seem small. While less can be said about the older stars, the sampling of 
sub-50-Myr systems is very complete within our distance bound. Also, our 
median distance for stars $\leq$50 Myr old is 75 pc, similar to 60 pc for 
older stars and 70 pc for systems with CO/C+ detections. We conclude that 
carbon-bearing gas appears in nearly half of the well-studied group of 
debris systems around A-stars that are up to 50 Myr old -- well beyond the 
sub-10~Myr epoch associated with gaseous proto-planetary discs.

The CO-hosting discs in Table 1 have dust luminosities characteristic of 
bright debris systems: $L_{dust}/L_{star}$ values range from 5 10$^{-4}$ 
to 7 10$^{-3}$ (Table 2), within the `luminous' classification of 
Mo\'{o}r et al. (2006). In contrast, proto-planetary discs are typified 
by $L_{dust}/L_{star} > 0.01$. This empirical boundary has recently been 
refined to criteria of R$_{12} < 3$ and R$_{70} < 2000$ (Wyatt et al. 
2015), where R is the ratio of dust-to-photospheric emission and the 
wavelength-subscripts are in microns. By this definition, HD 141569 may be an 
intermediate case, with R$_{12}$ of 6 ($L_{dust}/L_{star}$ of 0.007), 
but still fairly distinct from gas-rich Herbig Ae stars. Hales et al. 
(2014) present a discussion of recent CO detections in HAe-type discs.

\section{Discussion}

\subsection{Correlations}

%Figure 4 compares fluxes of dust continuum, OI, CII and CO (Table 2), 
%after adjusting all measurements to a common distance of 100 pc. The 
%five CO-hosting discs are ordered with increasing age (left to right) 
%in this barchart. The flux-proportions are fairly similar from system 
%to system, but radiative transfer modelling would be needed to compare 
%the relative abundances of dust and gases. There is little trend of 
%fluxes with age, except that OI drops below detection limits in older 
%systems as noted above.

\begin{figure}
\label{fig5}
\includegraphics[width=0.4\textwidth,angle=90]{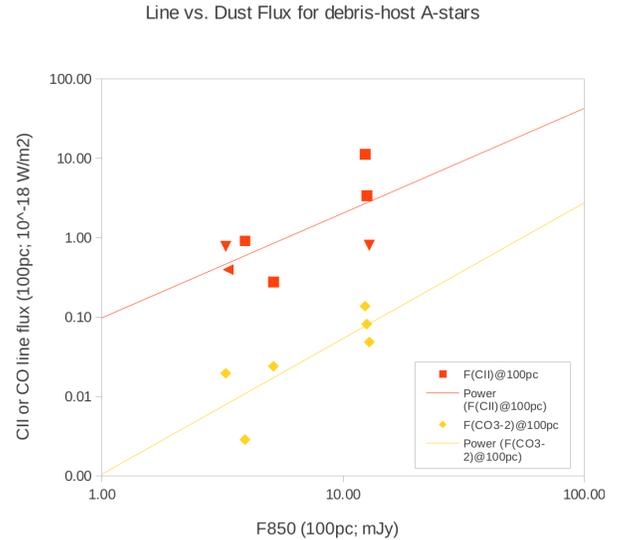}
\caption{Correlations of gas and dust line fluxes (Table 2, but corrected to 
a common distance of 100 pc). Triangles represent upper limits; 
$\eta$ Tel is omitted from the CO-correlation as it has limits on both 
axes (x $\leq$3.3, y $\leq$0.02).  
}
\end{figure}

\begin{figure}
\label{fig6}
\includegraphics[width=90mm,angle=0,trim=0 2cm 0 7cm,clip]{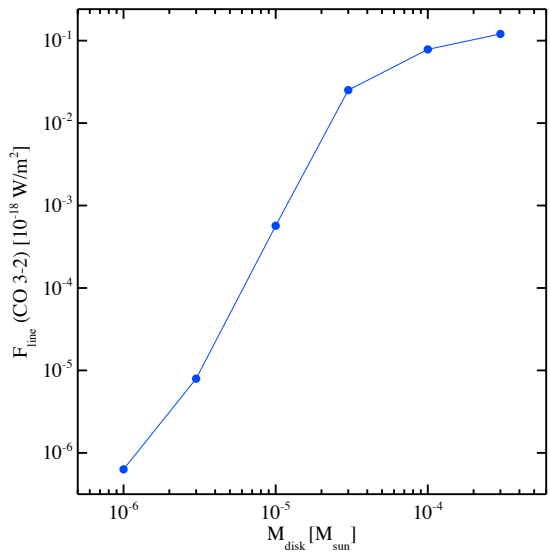}
\caption{ProDiMo code results. The 
host star is at 100 pc with T$_{eff}$ = 8600 K, L$_*$ = 14 
L$_{\odot}$, X-ray flux of 10$^{29}$ erg and UV flux of 0.1 L$_*$. The power-law 
distribution of dust sizes has a -3.5 index over 0.01 to 300 mm. The 
gas mass is ten times that of 
dust, and abundances are solar. The debris spans 30-100 AU in radius, 
with vertical height H(r) $\propto r^1$, to 10 AU height at 100 AU 
radius. The dust is settled to a thin mid-plane layer. The column density 
through the disc scales as $\Sigma(r) \propto r^{-1}$, with exponential 
tapering inward of 40 AU and outward of 50 AU.
}
\end{figure}

Figure 5 plots the CO and CII fluxes versus the 850 micron dust 
emission. There are only 6-7 data points to examine for any trends, but 
it appears that both gases rise in brightness more steeply than the 
dust flux. For CO, the data can be fitted as F(CO) $\propto$ 
F(860$\mu$m)$^{1.7}$. This is suggestive of a trend where the mass in 
CO may scale with the production rate (i.e as the square of the dust 
mass), but a more detailed treatment of gas excitation and opacity is 
needed (see Mo\'{o}r et al. (2015), Figure 8, for the results of such 
analysis). 

Here we have explored trends in CO flux versus disc mass, using the 
ProDiMo code (Woitke et al. 2016) to solve self-consistently for 
chemistry and radiative transfer in steady-state. The disc is assumed to 
have evolved into a `late' proto-planetary stage, with a gas-to-dust mass 
ratio of 10, and dust that has settled downwards to a thin plane. While 
the disc chemistry produces CO emission that rises more steeply than disc 
mass (Figure 6), the absolute values of the line flux are too small 
compared to those observed. In particular, dust masses (Mo\'{o}r et al. 
2015) of $\leq$1 M$_{\oplus}$ are at or below the left end of the x-axis, 
but CO J=3-2 line fluxes (Figure 5) lie at the top of the y-axis. This 
supports the hypothesis that these are {\it not} simply evolved 
proto-planetary discs, but that the CO molecules have a different origin. 
If their source is volatiles released from ices in collisions, then the 
starting point is not in fact an H-rich gas disc, and also the gas-phase 
chemistry will probably not be in steady state.

\begin{figure}
\label{fig7}
\includegraphics[width=0.5\textwidth,angle=0,trim=2cm 12cm 5cm 2cm,clip]{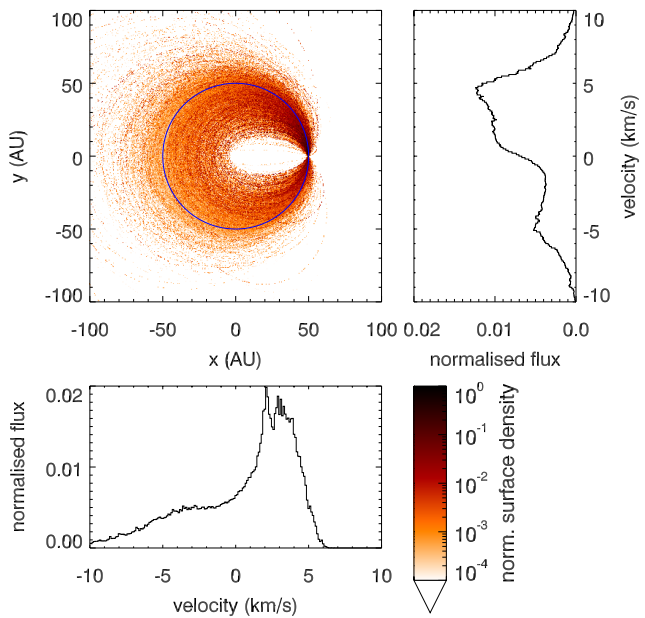}
\caption{
Example of CO production for a giant impact model.  The top-left panel
shows the spatial distribution of CO in a face-one view of the disk. 
The bottom and top-right panels show spectra for the CO as viewed
edge-on from the bottom and right of the image respectively. The impact
of an anti-clockwise-orbiting progenitor occurs at 50 AU from a 1.5 solar-mass 
star, producing a velocity dispersion of 0.3 times the circular Keplerian 
speed (see Jackson et al. 2014). The CO decays exponentially with a timescale 
of 120 years after production.
}
\end{figure}

\subsection{Asymmetries}

Models predict that CO can have a very asymmetric spatial distribution,
if the molecules originate from specific locations in the disc.  This
could occur in any model in which CO is produced by collisions between
CO-ice bearing bodies and there are inhomogeneities in the density (and
so collision rate) of the parent bodies. Examples of such models
include points resonant with a planet, or the site of the break-up of a
massive planetesimal.  The asymmetry in the debris is enhanced for CO
because of its short photodissociation timescale, so that the molecules
are most concentrated at their point of origin. Jackson et al. (2014)
have modelled this scenario for the case of a giant impact/break-up
event, and Figure 7 illustrates how the observational outcome can vary
for different viewing directions. The spectra are strongly asymmetric,
but the shape of the spectrum varies considerably with the (generally
unknown) viewing angle, complicating the interpretation.  An additional
complication is that the simple model in Figure 7 only includes the
Keplerian orbital motion of the CO gas, and does not include the effect
of gas pressure, which may be significant at higher CO densities.  This
would change the shape of the lines in Figure 7, but would not remove
the large scale asymmetry in the line profiles.

\begin{figure}
\label{fig8}
\includegraphics[width=0.5\textwidth,angle=0,trim=3cm 0 2cm 0,clip]{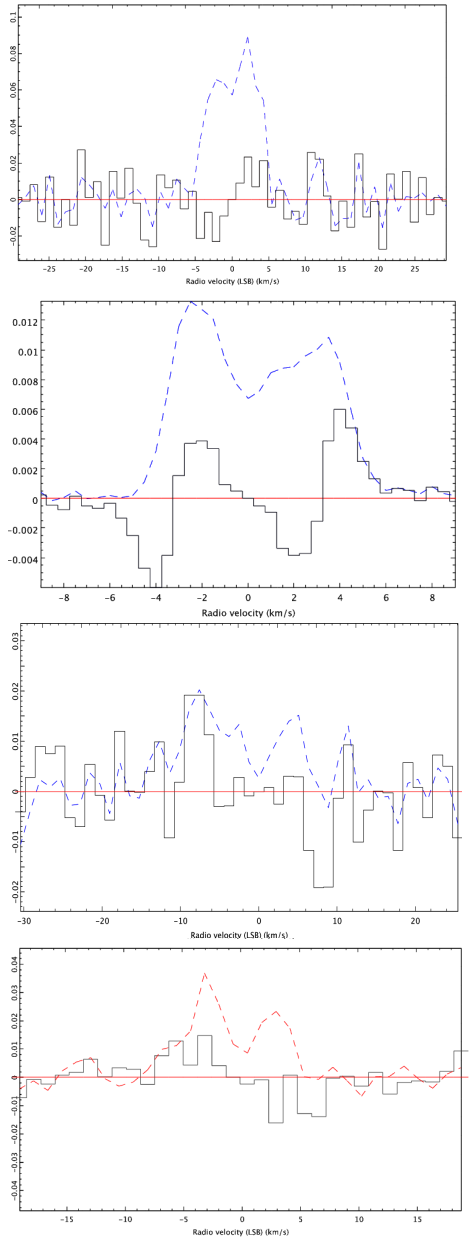}
\caption{CO whole-disc line profiles, and blue-red difference spectra (see 
text). From top to bottom: HD 141569 observed in CO 3-2; $\beta$ Pic 
in CO 2-1; HD 32297 in CO 2-1 (this work); 49 Cet in CO 3-2.
Spectra of HD 21997 and HD 131835 (from APEX: Mo\'{o}r et al. 2011, 2015) 
are symmetric within the noise, and are omitted. Y-axis units (not 
shown) are native to each dataset; actual (dashed) and difference spectra 
are on the same scale within each plot. X-axis units are velocity with 
respect to the line-centre minimum-signal point. 
}
\end{figure}

\begin{table}
 \caption{Asymmetry measures for gas and dust. For CO, the ratio given is 
of integrated intensity over two half-lines, with the mid-point defined as 
the faintest line-centre channel. For dust images, the ratio given is of 
integrated flux on either side of the star along the major axis; for the 
single dish data these estimates only compare two beams. Data include: 
HD 141569 -- CO 3-2 (JCMT), 0.87 mm dust (LABOCA); HD131835 -- CO 3-2 (APEX), 
0.87 mm (LABOCA); $\beta$ Pic -- CO 2-1 and 0.87 mm dust (ALMA); 
HD 21997 -- CO 2-1, 3-2 (APEX), 0.85 mm dust (JCMT); HD 32297 -- CO 2-1 
(JCMT), 1.3 mm dust (CARMA); 49 Cet -- CO 3-2 (JCMT, ALMA), 0.87 dust 
(ALMA). LABOCA images are from Nilsson et al. (2010); other references are 
given above. 
}
 \label{tab:asym}
 \begin{tabular}{@{}lccr}
  \hline
star		& CO line	& mm dust 	& notes	\\
		& asymmetry	& asymmetry	& 	\\
  \hline
HD 141569	& 1.0$\pm$0.2	& $\ga$1.5	& opt. thick CO	\\
HD 131835	& $\sim$1.3	& $\sim$1.3	& opt. thick CO?\\
$\beta$ Pic	& $\approx$2	& $\approx$1.2	& CO more asym.	\\
HD 32297	& 1.6$\pm$0.3	& $\approx$2	& similar asym.		\\
49 Cet		& $\sim$1.8	& $\approx$1.1	& CO more asym?\\
HD 21997	& $\approx$1.0	& $\approx$2.3	& opt. thick CO	\\
  \hline
 \end{tabular}
\end{table}

In Figure 8, whole-disc CO line-profiles are shown for four systems. For 
$\beta$ Pic, archival ALMA observations in CO J=2-1 were used to generate 
disc-integrated spectra (with sensitivity up to arcmin scales). For 49 
Ceti and HD 141569, we used CO 3-2 lines extracted from single-beam JCMT 
data\footnote{For HD 141569, data from 2009 were used, with earlier 
archival spectra from 1995 and 2001 showing similar profiles.}. The plots 
show tests for asymmetry, where the line velocities around the central 
dip have been swapped in sign, and used to generate a blue-red difference 
spectrum. Where this residual is non-zero, the emission from the the 
approach and receding sides of the disc is not the same, as for example 
in many sightlines across the model disc in Figure 7.

From this analysis, we find that CO asymmetry is most marked in $\beta$ 
Pic. The high signal-to-noise line is asymmetric in both height and width 
of the blue and red sides, reflecting the clump that has been spatially 
resolved (Dent et al. 2013). For 49 Ceti, the overall line shape appears 
similar to $\beta$ Pic, but the nature of any asymmetry is hard to 
characterise. Hughes (2014) has presented ALMA observations of the 49 Ceti 
disc (sampling scales only up to $\approx$10 arcsec) and the line profile 
shows some differences to that from the JCMT. Extended emission could thus 
be affecting the single dish spectrum, although it appears uncontaminated 
by the northern object seen by SCUBA-2 (ALMA sees no low-velocity CO 
here). The line asymmetry for HD 32297 is only at 2.5-sigma confidence, and 
the visual appearance of asymmetry for HD 141569 is at the level of noise 
in the difference spectrum.

Owing to the short dissociation time for CO molecules, whole-disc spectra 
might be expected to be more skewed than the dust distributions. Table 3 
attempts to quantify this using the integrated intensities of the blue/red 
sides of the CO lines, and the total dust fluxes observed to either side 
of the star. The available data are rather eclectic, but two systems do 
hint at the expected skewness behaviour. Both $\beta$ Pic and 49 Cet 
appear more asymmetric in CO than in dust, using these measures, although 
the CO profile for the latter is uncertain, as discussed above. HD 32297 
and HD 131835 have spectral data too noisy to discriminate, while HD 21997 
and HD 141569 appear more asymmetric in {\it dust}. However, these two 
systems appear to have optically thick CO lines such that saturation would 
maximise intensity in both line-halves. The less abundant isotopologue 
$^{13}$CO has recently been imaged in the 2-1 transition (HD 21997: 
Kosp\'{a}l et al. 2013; HD 141569: P\'{e}ricaud et al. 2016), and greater 
blue/red asymmetry begins to appear in these less saturated lines. Both of 
these studies found large masses of CO, however, that may be inconsistent 
with sources in a comet belt. The implications for the origin of the gas 
(collisional versus remnant proto-planetary discs) are discussed by 
Zuckerman \& Song (2012) and Mo\'{o}r et al. (2015).

\section{Conclusions}

With the discovery of CO molecules around HD 32297, there are now six 
A-type main-sequence stars where dusty debris is known to be accompanied 
by carbon monoxide gas. In the age bracket spanning 5-50 Myr, nearly half 
of the A-stars with debris in fact exhibit carbon-bearing gas. This 
suggests a prolonged active epoch where giant collisions release a burst 
of gas from frozen volatiles. The time period is similar to that taken to 
assemble the Earth from colliding planetesimals (e.g. Jacobson \& Walsh 
2015), so further study may yield clues to how volatiles are folded into 
rocky planets and their atmospheres.

As the photo-dissociation time for CO around A-type stars is short, the 
gas profiles will tend to be asymmetric if debris originates from one 
spatial location, with molecules mainly on one side of the star. This is 
supported for 1-2 systems where the gas asymmetry appears to exceed that 
of the dust. Models can explain a high incidence of CO detection when the 
molecules are short-lived if, for example, colliding debris repeatedly 
passes through the original impact point. However, at least two of the 
discs are optically thick in CO emission, suggesting the gas may not be 
collisional, but instead represent a very prolonged proto-planetary disc 
phase.

\section*{Acknowledgments}

Data were obtained under JCMT project IDs MJLSD01 and M13BU16. JSG and PW 
thank the ERC for funding for project DiscAnalysis, under the grant 
FP7-SPACE-2011 collaborative project 284405. JPM is supported by a UNSW 
Vice-Chancellor's postdoctoral fellowship. MCW and LM acknowledge the 
support of the European Union through ERC grant 279973. 

The James Clerk Maxwell Telescope is operated by the East Asian 
Observatory on behalf of The National Astronomical Observatory of Japan, 
Academia Sinica Institute of Astronomy and Astrophysics, the Korea 
Astronomy and Space Science Institute, the National Astronomical 
Observatories of China and the Chinese Academy of Sciences (Grant No. 
XDB09000000), with additional funding support from the Science and 
Technology Facilities Council of the United Kingdom and participating 
universities in the United Kingdom and Canada. ALMA is a partnership of 
ESO (representing its member states), NSF (USA) and NINS (Japan), together 
with NRC (Canada), NSC and ASIAA (Taiwan), and KASI (Republic of Korea), 
in cooperation with the Republic of Chile. The Joint ALMA Observatory is 
operated by ESO, AUI/NRAO and NAOJ.

\bsp

\label{lastpage}

\end{document}